# Stability Mechanisms of High Current Transport in Iron-Chalcogenides Superconducting Films

A. Leo, G. Grimaldi, P. Marra, R. Citro, F. Avitabile, A. Guarino, E. Bellingeri, S. Kawale, C. Ferdeghini, A. Nigro, S. Pace

*Abstract*— The improvement in the fabrication techniques of iron-based superconductors have made these materials real competitors of high temperature superconductors and MgB$_2$. In particular, iron-chalcogenides have proved to be the most promising for the realization of high current carrying tapes. But their use on a large scale cannot be achieved without the understanding of the current stability mechanisms in these compounds. Indeed, we have recently observed the presence of flux flow instabilities features in Fe(Se,Te) thin films grown on CaF$_2$. Here we present the results of current-voltage characterizations at different temperatures and applied magnetic fields on Fe(Se,Te) microbridges grown on CaF$_2$. These results will be analyzed from the point of view of the most validated models with the aim to identify the nature of the flux flow instabilities features (i.e., thermal or electronic), in order to give a further advance to the high current carrying capability of iron-chalcogenide superconductors.

*Index Terms*— Superconducting thin films, iron based superconductors, transport properties, vortex dynamics

## I. INTRODUCTION

THERE ARE different parameters which set limits to the feasibility of using a particular superconducting material in high power applications [1]. Among these parameters, the critical current density $J_C$ is one of the most important since it is the maximum current density a superconductor can carry without dissipation. On the other hand, in some practical applications like phonon detectors or fault current limiters, the knowledge of the material behavior in the dissipative regime above $J_C$, usually referred as flux flow regime, is crucial. Indeed, in these applications, a rapid switch from the superconducting regime to the normal state is appreciated: a circumstance far opposite to the case of the high power applications, where a more gradual transition allows a greater quench control. Thus, it is natural that the study of the phenomena known as flux flow instabilities (FFI) has attracted a great interest in the past years and it is still an open research field [2-8].

The FFI manifests itself as a sudden jump from the flux flow branch to the normal state in current driven current-voltage characteristics. Such feature could be seen as detrimental if aiming to a highly stable current carrying superconductor or it can be a way to boost the performances of those devices based on the superconducting-to-normal state switching. Moreover, when triggered by intrinsic mechanisms, the FFI is a valuable tool for investigating quasiparticle scattering processes, especially when the quasiparticle distribution is far from equilibrium. Indeed, the presence of the FFI is related to the loss of stability in the electric current transport into the superconductor due to a dramatic change in the moving vortex structure. Regardless of the mechanism leading to it, the change in the vortex structure is always caused by an imbalance between the quasiparticles energy relaxation processes, mainly scattering and recombination [9,10].

Nowadays, many efforts have been made to improve the current capability of superconducting wires and tapes based on iron-based superconductors (IBSC). In particular, for the iron-chalcogenide superconductors we mention the successful realization of coated conductors able to carry very high current densities ($J_C$ at 4.2 K and 30 T up to $10^5$ A cm$^{-2}$) [11] and the identification of a fabrication technique able to maximize the critical current density value with the proper choice of the substrate [12-14]. Instead, the current carrying stability of IBSC in the flux flow regime is still an almost unexplored field.

In the present paper, we study the stability of electric current transport in Fe(Se,Te) microbridges from optimized thin films in a regime well above the critical current. Quench features similar to those related to FFI have been recognized in this type of samples and the contributions to the FFI from extrinsic thermal mechanisms, as well as from intrinsic electronic mechanisms, have been investigated. Comparing our data with both the experiments and the theoretical predictions from the literature, we will enlighten which are the mechanisms contributing to the observed quench phenomenon.

. The research leading to these results has received funding from the PON Ricerca e Competitività 2007-2013 under grant agreement PON NAFASSY, PONa3_00007. *(All authors contributed equally to this work.) (Corresponding author: Antonio Leo.)*

A. Leo, A. Guarino, R. Citro, A. Nigro, and S. Pace are with Physics Department 'E. R. Caianiello', Salerno University, via Giovanni Paolo II, 132, Stecca 9, I-84084 Fisciano (SA), Italy. They are also with CNR-SPIN Salerno, via Giovanni Paolo II, 132, Stecca 9, I-84084 Fisciano (SA), Italy, e-mail: antoleo@sa.infn.it; guarino@sa.infn.it; citro@sa.infn.it; nigro@sa.infn.it; pace@sa.infn.it.

F. Avitabile is with Physics Department 'E. R. Caianiello', Salerno University, via Giovanni Paolo II, 132, Stecca 9, I-84084 Fisciano (SA), Italy, e-mail: favitabile@unisa.it.

G. Grimaldi and P. Marra are with CNR-SPIN Salerno, via Giovanni Paolo II, 132, Stecca 9, I-84084 Fisciano (SA), Italy, e-mail: gaia.grimaldi@spin.cnr.it; pasquale.marra@spin.cnr.it.

E. Bellingeri, S. Kawale, and C. Ferdeghini are with CNR-SPIN Genova, corso Perrone 24, I-16152 Genova, Italy, e-mail: emilio.bellingeri@spin.cnr.it; shrikant.kawale@spin.cnr.it; carlo.ferdeghini@spin.cnr.it.

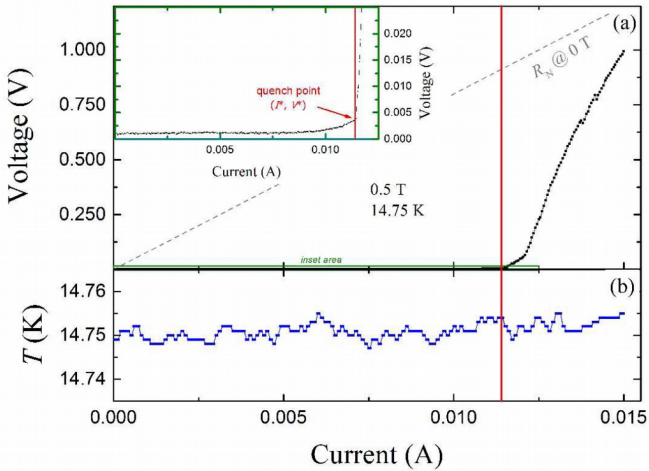

Fig. 1 (a) An example of Current-Voltage Characteristic acquired at a temperature of 14.75 K and in an applied magnetic field of 0.5 T. Inset: an enlargement of the low voltage data shown in the main panel. (b) The sample holder temperature variation as a function of the bias current. The vertical red line in the panels marks the point at which the quench occurs.

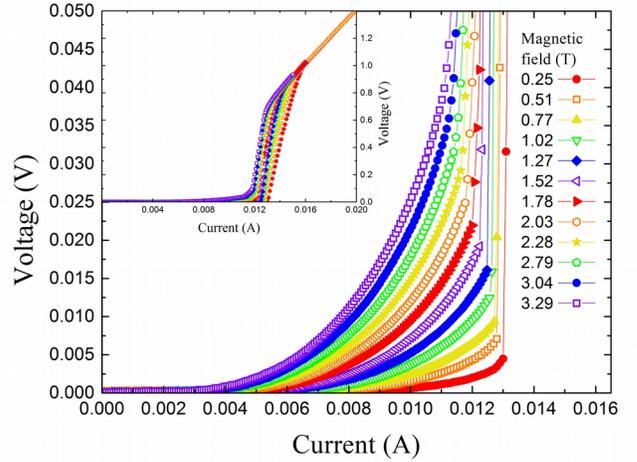

Fig. 2. Enlargement of low voltage region of Current-Voltage Characteristics acquired at a constant temperature of 16.00 K and in different applied magnetic fields. In the inset: the CVC's in the whole measured range.

## II. EXPERIMENTAL DETAILS

### A. Samples description

The samples considered for the present work are microbridges obtained by standard UV photolithography and Ar ion-milling etching form Fe(Se,Te) thin films grown on a $CaF_2$ substrate. The films have been fabricated by an optimized Pulsed Laser Deposition (PLD) starting from a target with the nominal composition $FeSe_{0.5}Te_{0.5}$. The resulting film thickness $d$ is about 120 nm, while the width $W$ of the microbridges is 20 μm and the distance $L$ between the voltage tips is 65 μm. The typical critical temperature $T_C$ of the samples is about 20.5 K, higher than the typical values of bulks samples. This increase on the $T_C$ is unambiguously related to the shrinkage in the $a$-axis induced by the chosen substrate [12]. Energy Dispersive Spectroscopy analysis preformed on a wide area, with lateral dimensions of some hundreds of micrometers, showed an high homogeneity of the films and revealed the following stoichiometry: $Fe_{0.98}Se_{0.67}Te_{0.33}$. More information about sample fabrication and their structural and pinning properties can be found elsewhere [12, 15].

### B. Measurement setup

Current-Voltage Characteristics (CVC's) have been acquired in a Cryogenic Ltd. cryogen free cryostat equipped with an integrated cryogen-free variable-temperature insert operating in the range 1.6–300 K and a superconducting magnet able to generate a field up to 16 T. These measurements have been performed by a pulsed current 4-probe technique, with a SourceMeter Keithley model 2430 used both as current source and voltage meter. The temperature has been measured with a Lake Shore Temperature Controller model 350 connected to a Lake Shore Cernox sensor model CX-1030-SD-1.4L mounted on the same metallic block used as sample holder. The sample is cooled by a continuous Helium gas flow and temperature stability is within 0.01 K.

In our CVC measurement, each current pulse has a rectangular shape with a power-on time (or Pulse Width, PW) equal to 2.5 ms; the time separation between each pulse (or Pulse Delay, PD) is set to 1 s in order to allow complete recover of the sample temperature to the He flow temperature. The sample holder temperature $T$ is monitored during the whole CVC acquisition; $T$ values are acquired just before each current pulse.

## III. CURRENT STABILITY ANALYSIS

In Fig. 1 we report one of the measured CVC's showing a clear evidence of the quench from the flux flow branch toward the normal state. The quench point is identified by the current-voltage values ($I^*$, $V^*$), as reported in the inset of the above mentioned figure. We also note that all the measured CVC's do not show any hysteresis.

The behavior exhibited by the CVC's at a fixed temperature and in different applied magnetic field reported in Fig. 2 induces to interpret such a rapid transition as a current instability. Indeed, looking at the dependence of the so-called instability current $I^*$ on the magnetic field reported in Fig. 3(a), we can recognize how $I^*$ is almost constant. On the contrary, the critical current, defined by a standard 1 μV/cm criterion, rapidly decreases as a function of the perpendicular applied magnetic field. Moreover, as shown in Fig. 3(b), the instability current behavior keeps being the same as the temperature increases, although the pinning as expected decreases. Usually, this feature is observed in high-$T_C$ superconductors [16], but not in low-$T_C$ materials [17]. Such a plateau of the instability current as a function of the applied field can be interpreted as a robust wide range of current stability above the critical current, similarly to what happens in superconducting materials modified with a proper pinning landscape [18,19].

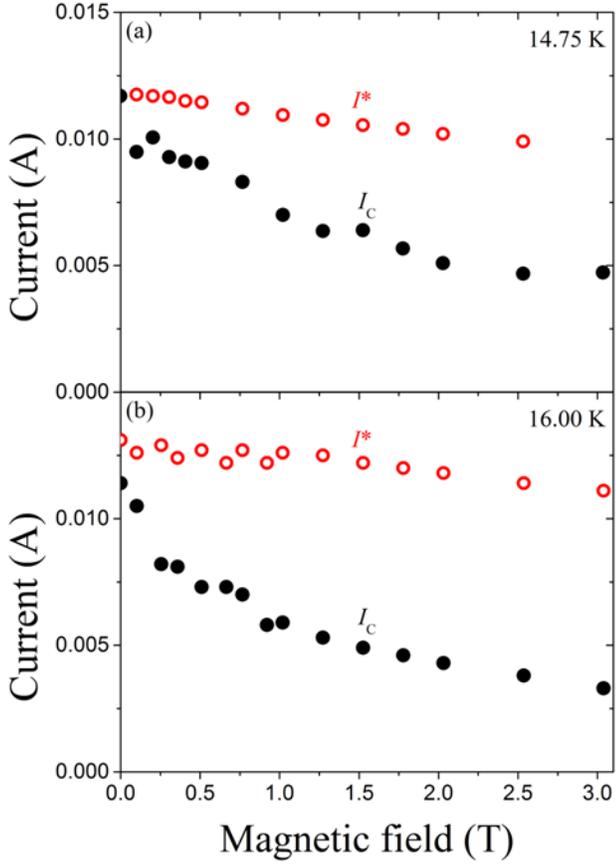

Fig. 3. Critical current $I_C$, defined by a standard 1 μV/cm criterion, and instability current $I^*$, defined as the current corresponding to the quench point in the CVC, as a function of the applied magnetic field for (a) $T$ = 14.75 K and (b) $T$ = 16.00 K.

This peculiar behavior of $I^*$ has been reported as a feature related to the presence of Larkin-Ovchinnikov (LO) electronic instabilities in high-$T_C$ superconductors [3]. Moreover, for the critical vortex velocity $v^*$ as a function of the applied magnetic field, with $v^*$ evaluated from the measured critical voltage curve $V^*$ since $v^* = V^*/(\mu_0 H \cdot L)$ [20], we can recognize the power law behavior $\mu_0 H^{-1/2}$ which is commonly associated with the presence of FFI triggered by intrinsic electronic mechanisms [16]. On the other hand, the transition from the flux flow branch to the normal state in our case is not that sudden jump usually observed when pure electronic mechanisms trigger the FFI. Thus, it is worth to evaluate the possible presence of a thermal contribution to the observed instability.

In the lower panel of Fig. 1, the variation of the temperature $T$ with the increasing bias current $I$ is shown. It can be clearly seen how the temperature excursion does not exceed the stability range, thus a complete recover of the sample holder temperature to the flow temperature during the PD is achieved. The absence of any hysteresis in the CVC's is also symptomatic of a relatively small increase, if present, of the sample temperature $T'$ during the quench, a circumstance which could be ascribed to the proper choice of the employed bias mode.

Even if a direct measurement of the increase of $T'$ is inaccessible, it is still possible to evaluate it by an indirect numerical method. This fitting procedure has been extensively reported elsewhere [20]. Once the fitting parameter are known, it is possible to reconstruct the actual temperature behavior of the film in order to evaluate the temperature increase at the quench point. In our case, it results a $\Delta T'$ of about 0.3 K, a value too low to ascribe such a rapid quench to only thermal mechanisms.

There is another parameter whose evaluation can add information about the contribution of pure Joule self-heating on the observed current instabilities. Indeed, following a microscopic analysis of the heat removal from a thin film in the resistive state near $T_C$, Bezuglyj and Shklovskij (BS) introduced a macroscopic parameter $B_T$ which separates the region where non-thermal intrinsic ($B \ll B_T$) or pure heating extrinsic mechanisms ($B \gg B_T$) of the instability dominates [21]. This parameter is expressed in terms of the quasi-particle energy relaxation time $\tau_E$ and the heat transfer coefficient $h$ as:

$$B_T = \frac{0.374 \cdot e \cdot h \cdot \tau_E}{k_B}\left(R_N \frac{W}{L}\right) \quad (1)$$

with $k_B$ the Boltzmann constant, $e$ the electron charge and $R_N$ the normal state resistance, which in our case is $R_N = 80\ \Omega$.

We have estimated $\tau_E$ using the LO expression [2], while $h$ is evaluated from the curve of the Joule power $P = V \cdot I$ as a function of the sample temperature increase $\Delta T'$ [20]. The estimated values are $\tau_E = 6.5 \cdot 10^{-10}$ s and $h = 30$ W·cm$^{-2}$·K$^{-1}$, consequently the evaluated $B_T$ is about 28 T. Since in our measurements the field range is well below 28 T, this is a further indication that the contribution from thermal mechanisms to the observed instability is not the determining one.

IV. CONCLUSIONS

Summarizing, quench features similar to those related to flux flow instabilities have been recognized in optimized Fe(Se,Te) thin films grown by Plasma Laser Deposition on a CaF$_2$ substrate. The nature of the mechanisms leading to this FFI has been investigated.

On the basis of the results reported in the present work, we can argue that both intrinsic and extrinsic mechanisms play a role in the observed quench phenomenon. Indeed, our results enlighten the existence of many points of contact between our experimental findings and the literature related to flux flow instabilities triggered by intrinsic electronic mechanisms in high-$T_C$ superconductors. On the other hand, the estimation of the sample temperature increase and the calculated value of the Bezuglyj-Shklovskij parameter identify the contribution from extrinsic thermal mechanisms as of a secondary relevance. Finally the current transport has been found to be intrinsically stable, even as a function of the magnetic field, thus confirming the high-field performance of these materials for applications.